# William Sadler Franks and the Brockhurst Observatory

**Jeremy Shears**

**Abstract**

William Sadler Franks (1851-1935) was astronomer-in-charge at F.J. Hanbury's private observatory at Brockhurst, near East Grinstead, Sussex, from 1909 until his death in 1935. This paper reviews the observational projects Franks undertook at Brockhurst, including his work on double stars, red stars, diffuse nebulae and dark nebulae, as well as his involvement with the commissioning of a 24-inch reflector built by Thomas William Bush (1839-1928).

## Introduction

During 2013 Richard Baum kindly presented me with three notebooks of astronomical observations made by William Sadler Franks at the Brockhurst Observatory, near East Grinstead, Sussex, between 1909 and Franks's death in 1935. They mainly contain notes of his observations of double stars, the moon, planets, nebulae and any passing comets that he made with the observatory's 6$^1$/$_8$-inch (15.5 cm) Cooke refractor. Richard had looked after the notebooks since they were given to him in about 1956 by Patrick Moore (1923-2012). Patrick Moore, as a boy in the 1930's, had known Franks and was a regular visitor to the Brockhurst Observatory, which lay across the road from the Moore family home. The observatory was owned by the wealthy industrialist F.J. Hanbury (1851-1938) and Franks, as the astronomer-in-charge, was expected to entertain Hanbury's guests by showing them celestial objects through the telescope. For the rest of the time, Franks was free to use the observatory as he wished. When Franks died in 1935, Hanbury invited the young Patrick Moore to take on the same function. This arrangement came to an end when Hanbury himself died in 1938 and the following year the observatory was sold. Franks's notebooks were handed to Patrick Moore, along with his personal copy of *Webb's Celestial Objects*, the sixth edition of which he has helped to prepare for publication. Over the years Patrick Moore passed several of the notebooks on to others, but it is hoped one day to reunite them as part of the BAA archive. Patrick Moore described his personal recollections of his involvement with Franks and the Brockhurst Observatory in a paper in this *Journal* in 2002 (1). Franks was a major influence on the young Patrick Moore.

Reading the three notebooks now in my possession provides an insight into Franks's astronomical interests and how they evolved over the years. What soon becomes clear is that Franks was, in modern parlance, "project orientated". That is to say, he tended to concentrate much of his observational work on a specific project, such as double stars, diffuse nebulae, dark nebulae, for a particular period, usually concluding with the publication of a paper on the subject, before moving onto a new project. Of course, he also kept an eye on other comings and goings in the sky, such as planetary oppositions and bright comets. Franks also made notes of other non-astronomical happenings, such as meteorological events, and which visitors he entertained at the observatory. He also describes how he maintained and improved the observatory's instrumentation. Of special note is his description of the installation and testing of a large reflector at Brockhurst: the 24-inch (60 cm) Newtonian which was built by Thomas William Bush (1839-1928).

This paper explores Franks's observational work at Brockhurst as well as the commissioning of the Bush telescope, based on his notebook writings.

## W.S. Franks's early life

Franks was born at Newark, Nottinghamshire, on 1851 April 26 and was initially involved in his father's business in Leicester (2). He became interested in astronomy at an early age and soon set up a small observatory. He developed a particular interest in star colours, which he maintained throughout the rest of his life. His first major publication was *A Catalogue of the Colours of 3890 Stars* which was communicated to the RAS on his behalf by the Reverend T.W. Webb (1807-1885) in 1878. It was therefore no surprise when Franks became Director of the Star Colours Section of the Liverpool Astronomical Society in the 1880s and subsequently served in the same capacity in the BAA from its establishment in 1890 until 1894 when he handed the baton to G.F. Chambers (1841-1915).

In 1892, Franks joined Isaac Roberts (1829-1904) at his Crowborough observatory where he was engaged in photographing nebulae and star clusters with the 20-inch (50 cm) reflector. Many of the impressive photographs presented by Roberts at meetings of the RAS to warm applause were taken by Franks. Upon Roberts's death in 1904, he spent a further two years assisting Roberts's widow, Dorothea Klumpke Roberts (1861-1942), in organising her late husband's records and closing the observatory. Having completed that task in 1906, Franks moved to Uxbridge during which time he had several small engagements connected with private observatories, including John Franklin-Adams's Mervel Hill Observatory, near Godalming in Surrey. Franklin-Adams (1843-1912; Figure 1) had carried out a photographic charting programme of the southern sky from South Africa and Franks assisted him in preparing the photographs for publication. Franklin-Adams was preparing to return to the Cape in 1909 when he passed away, leaving Franks's future again uncertain.

## The move to Brockhurst

Franks's involvement with the Brockhurst Observatory began in September 1909. The observatory was located on the Brockhurst estate of Frederick Janson Hanbury, whose personal wealth was established through the family firm of Allen and Hanbury Ltd, a pharmaceutical company, of which he was senior partner (3). Hanbury's main interest was in botany and horticulture; he was a well-known expert on orchids. He maintained a large staff of gardeners and turned Brockhurst into a world-famous garden in the 20 years following his purchase of the estate in 1908 (4). The observatory had a painted canvas dome (Figure 2) which housed the $6^1/_8$-inch f/13.3 Cooke refractor on a mahogany English equatorial mounting. In addition there was a portable tripod-mounted 3½-inch f/13.7 refractor. It appears that Hanbury was no great observer himself, but he enjoyed entertaining his guests, about whom we shall hear more later, by showing them objects after dinner and this is where Franks came in as "observer" (5).

Franks's first notebook, *Astronomical Diary and Observations (I)* (6), covering the period 1909-23 (Figure 3) begins with a description of the observatory written during his first visit after his appointment, on 1909 September 11. Franks's initial task was to organise the remounting of the Cooke refractor, which involved raising the mount's piers and the roof a few inches. He does not explain why this was necessary, but it is possible that a smaller telescope had been installed previously. At the same time adjustments were made to the driving clock mechanism on the RA gear and taller observatory steps were ordered. Franks's first observations with the remounted telescope were of the sun and stars on September 23. For the first few months he was still living in Uxbridge, so most of his visits were limited to daytime and early evening. He occasionally recorded observations made at Uxbridge including his first sighting of comet 1P/Halley on 1910 April 26, which he also followed throughout May from Brockhurst. By the autumn of 1910 he was living near East Grinstead, within cycling distance of the observatory, so he was able to intensify his observing programme. He named his house "Starfield" after Isaac Robert's observatory. Nevertheless he made two extended trips back to Mervel Hill during October to conclude his work on the late Franklin-Adams's observations. Franklin-Adams's *Photographic Chart of the Heavens* was eventually published between 1912 and 1914.

As well as astronomical observations, Franks also initiated a programme of meteorological recording in the summer of 1911. However, he was shocked by an act of apparent vandalism on his new weather station on August 18: "Found, on arrival at Brockhurst this morning, that a wanton outrage had been committed during the night by some mischievous intruders. All the meterol[ogical] instr[uments] were deliberately smashed; all woodwork injured and even the brick pillars knocked over bodily. The glass ball of [the sun] recorder was missing – either hidden or stolen". He therefore had to re-establish the station and it wasn't until 1913 January that the Campbell-Stokes sunshine recorder was used again, for the glass ball of was recovered from a nearby hedgerow, largely undamaged, save a slight chip (7).

**Visitors to the Brockhurst Observatory**

Judging by the accounts in Frank's notebooks, entertaining visitors was not a particularly onerous task. Typically only one or two groups of visitors would descend upon him a month and often several months went by without any. The organisation which attended most regularly was the local troop of Girl Guides who were generally treated to two evenings at Brockhurst every August. At other times local church groups and YMCA members attended. But of course it was Hanbury's personal dinner guests that were really that ones that Franks needed to take care of. Many of these were fellow plant experts or gardeners, but business associates and national dignitaries were also represented. Probably the most illustrious visitor was HRH the Crown Prince Gustav Adolf of Sweden (1882-1973; Figure 4), the future King Gustav VI Adolf (8), who visited on 1925 August 13, along with his aide-de-camp. The Prince was also a renowned horticulturalist, known for his extensive rhododendron collection. Other dignitaries included Lord and Lady Aberconway (9), who developed the gardens at Bodnant in north Wales, Sir Arthur Fenton Hort, a plantsman (10), Sir George Truscott, local Baronet and businessman (11), and Lady Lockyer, the widow of the spectroscopist Sir Norman Lockyer (12).

Of course, it was the visitors that Franks hosted on the evening of 1933 August 21, a Monday, that were particularly auspicious in an astronomical context: a certain "Mr. Moore and family". They must have been impressed with their view of Vega, the "double-double star" ε Lyr and the Ring Nebula, M57, for "Mr. Moore and son", the 10-year old Patrick, returned to the observatory the following week and "looked at a number of interesting objects" through the telescope (13).

**Franks's first observations at Brockhurst**

During 1910 and 1911, Franks undertook observations of a wide variety of objects, including double stars, coloured stars (14), nebulae and planets. He had an usually clear view of Venus at half-phase on 1911 December 6, when "[b]esides the normal fading off towards the term[inator], felt quite certain of some ill-defined dusky patches (something like lunar seas, only greatly attenuated), though they are not definitive enough to draw". He reported a similar effect in 1913 (15): "disc distinctly mottled and patchy – there seemed something like lunar 'seas', but so v. faint as to defy delineation".

Franks was also afforded some splendid views of Mars during the 1911 opposition. One of his best views was on December 13, after a dull day with some rain, when he "Got 2 sketches of Mars…def[inition] v. good…..the details came out unusually well – the planet having more of a map aspect than usual". These sketches can be seen in Figure 5.

One of the first projects that Franks engaged in was to investigate the limiting magnitude of stars that could be observed during daylight. With the $6^1/_8$-inch refractor he could certainly see down to mag 4.6, providing that the sky was free of haze, the star was more than 60° from the sun and the horizon and within 3 hours of the meridian (16). He found a magnification of x200 to be best, although the colours of the stars were not as vivid at night. He found 187 daylight stars in this way.

Observing conditions at Brockhurst were excellent, with the zodiacal light being visible regularly. It wasn't until 1931 that the nearby main Lewes Road was illuminated with electric lights (17).

**Double star measurements**

In 1913 Franks set about a programme of improving the observatory facilities to enable him to embark on a project of double star observations, which was to consume much of his observing time over the next seven years. Although he had access to a micrometer he found the results unsatisfactory, thus during the summer he ordered a new micrometer from T. Cooke & Sons of York. At the same time he oversaw the construction of a transit room adjoining the main observatory (this can be seen in Figure 2), which meant that observing had to be suspended during May and June. The transit instrument, a 2¾-inch (7 cm) f/11.6 refractor by Troughton & Simms, was duly installed and Franks carefully sketched the view of the meridian horizon looking south towards Forest Hills, about 5 miles (8 km) away, including the positions of the trees (Figure 6).

Measurements with the $6^1/_8$-inch refractor equipped with the new micrometer began in earnest in September. By the end of the first month Franks had made measurements of 73 stars on 21 nights and by the end of the first year of the project he had 838 measurements of 575 stars on 164 nights. The first paper resulting from the project was published in 1914 April, entitled "*Micrometrical measures of 360 wide double stars*" (18), followed by another in December (19), with at least one additional paper appearing each year until 1920 (20). In recognition of his work, in 1923 he was awarded the Jackson-Gwilt medal of the RAS (21). The high quality of his work quickly became known widely and he received a specific request from R.G. Aitken (1864-1951) of the Lick Observatory to provide measurements of large proper motion double stars (22).

During the first year of the project Franks found that he needed to make several adjustments to the telescope mount to improve its tracking accuracy and thereby facilitate the use of the micrometer. This involved replacing the RA bearings and installing a new drive gear, which is shown in Figure 7. He was particularly proud of the fact that the latter was completely designed and fabricated in the Brockhurst estate workshops by one of the machinists, Mr. J.S. Adams (23). After several months of effort in fine-tuning the equipment, Franks heralded his opinion that the instrument "is now in first class order; probably even better than when it first left the maker's hands many years ago" (24). Even then, some double stars were too challenging for the Cooke telescope to resolve and occasionally Franks would seek permission to use the 28-inch (71 cm) refractor at Greenwich (25).

The pace and focus with which Franks worked on the double star project is remarkable. Apart from occasional equipment problems – the micrometer was sometimes sent to T. Cooke & Sons for adjustment or to have its hairs replaced - and maintenance breaks (26), observations were made virtually every clear night. The only interruptions to his work were the visitors and on one occasion he wrote tersely in his notebook: "A really splendid night, but unfortunately best part of it was monopolised by visitors" (27). Another annoyance was the introduction of "Daylight Savings" time in World War One, which meant it didn't get dark till late, a situation which Franks found "an unmitigated nuisance" as it reduced the number of measurements he was able to make during the summer months (28). The Great War received little other comment from Franks, although he did wonder whether the torrential rainfall experienced during 1917 July might have been connect with the "unprecedented artillery duel in Flanders" (29) - a potential link much debated in the columns of the *English Mechanic* at the time. He also heard an explosion during the evening of 1917 May 20 during a thunderstorm, which turned out to be the result of lightning striking a munitions works at Tonbridge, Kent, some 25 km away, causing it to explode (30).

**Hagen's and Espin's red stars**

Franks does not explain what brought his double star observations to an end, but it appears that he made relatively few observations during the period between submitting his final double star list to the RAS in 1920 and early 1923 when he embarked on a new project. This involved returning to his first love: the observations of the star colours. The renewed activity was apparently stimulated by a request from Father J.G. Hagen (1847-1930) of the Vatican Observatory to determine the mean visual colour of around 6000 red stars (31). Franks soon extended the work to orange and red stars in the Draper Catalogue (32), which he presented to the RAS on 1924 November 14, and, in 1925, to observations of red stars recently listed by T.H.E.C. Espin (1858-1934) (33). Franks took a break from his red star observations during late summer and autumn 1924 to observe Mars at opposition, during which he was able to see the satellites Phobos and Deimos for the first time in his life, over several nights.

Then in 1925, Franks became involved with commissioning the 24-inch Bush reflector at Brockhurst and it is to that we shall now turn our attention. However, we shall first consider who "Bush" was.

**Thomas William Bush: a short biography**

Like Franks, Thomas William Bush (1839-1928) was also a native of Nottinghamshire, being born at Nottingham on 1839 May 19 (34). His father was a dyer and he passed away when Bush was seven years old. His mother remarried John Marriott, a baker and flour seller in the city. In spite of his humble surroundings, Bush evidently did well at school and studied German, Greek, Hebrew and Latin in his own time. In his early twenties he joined the Nottingham Mechanics Institute and whilst there his astronomical interest was kindled though attending two lectures by the selenographer W.R. Birt (1804-1881). He began to grind mirrors and to build telescopes. Over a period of seven years he built a 13-inch (33 cm) reflector which was exhibited at the 1870 Working Men's International Exhibition in London, where it was awarded a gold medal (Figure 7) (35). Both Queen Victoria and Prime Minister W.E. Gladstone (1809-1898) viewed the instrument. Gladstone was suitably impressed with the workmanship for he informed G.B. Airy (1801-1892), who, wishing to encourage Bush, sent him a Browning spectroscope, a solar filter and a micrometer. This spurred him on to even greater efforts in telescope building.

At the same time, Bush's professional life and status blossomed. Having followed his step-father into the baking trade, he move on in professional circles and was appointed as Secretary of Nottingham General Hospital in 1873, the same year he was elected FRAS. Selling his bakery and grocery business, he raised sufficient funds to build a comfortable three-bedroom house for his wife (36) and himself at Thyra Grove, Mapperley, Nottingham. In the grounds he constructed an observatory for his 13-inch telescope, with an adjoining transit room. The Mapperley Observatory, as it was known, opened in the spring of 1877. His astronomical work was becoming well-known amongst the burghers of Nottingham, who presented him with a sidereal clock in 1879.

Orlando Weld-Forester (1813-1894), was rector of Gedling parish church near Nottingham from 1867 until his elevation to the peerage, as Lord Forester, in 1887. He was a keen amateur astronomer and Bush got to know him well during his time at Gedling. In 1889, Bush resigned his position at Nottingham Hospital and moved to Lord Forester's observatory at Willey Park where he was placed in charge of the 8-inch (20 cm) refractor. Lord Forester died in 1894, but Bush remained for several more years, helping to run the estate.

Finally, at the age of 70, in 1909, Bush moved to East Grinstead where he continued to build several large telescopes, including the 24-inch reflector that was erected at Brockhurst. His vision was to construct photographic telescopes building the mechanical parts and grinding the mirrors himself. One 24-inch telescope was destined for Nottingham University and another for Reading University (37). At East Grinstead he got to know W.S. Franks well. When Bush died on 1928 April 28, his RAS obituary was written by Franks (38).

**The 24-inch Bush reflector**

With advancing years, and by now a widower, Bush moved from his home into Sackville College, a residential home for elderly gentlemen on East Grinstead High Street on 1924 January 8 (39). As a result it was decided to install Bush's 24-inch telescope at Brockhurst under Franks's supervision. Franks's notebooks describe the installation and testing of the telescope in some detail.

Housing the Bush telescope was clearly a major challenge due to its sheer size, which also necessitated ordering new steps, 6½-feet (2 m) high to access the eyepiece at the Newtonian focus (40). Thus a run-off shed was erected at Brockhurst during the summer of 1925. It was a corrugated iron structure, which was essentially a movable Nissen hut, some 4.3 m long, 3.2 m wide and 2.7 m high at the top of the corrugated arch. The structure moved on rails twice this length. There were York stone foundations, brick piers for the mount and an elevated boarded floor. Views of the telescope can be seen in Figure 9.

Whilst the observatory was ready by the end of 1925 September, installing the telescope and the electrical fittings took some months more, which meant that Franks wasn't able to achieve "first light" until the following May and even this was only with an unsilvered mirror, with the aim of performing some basic tests and determining the focal length accurately. The telescope "tube" was square in cross-section, being fabricated from an iron framework enclosed with wood panels. It was mounted on an equatorial fork mount with a weight-driven RA drive constructed from brass. The instrument was equipped with an 8½-inch (22 cm) guide scope (41), to facilitate guiding during long-exposure photography, although the instrument was never used for this purpose, and a 2-inch (5 cm) finder scope.

Bush had initially supplied two 24-inch mirrors, which had an effective aperture of 23¾-inches taking into account the retaining ring:

Number 1: 1½-inch thick, focus = 98-inches (f/4.1), with a 5-inch central hole

Number 2: $2^3/_8$-inch thick, focus = 96-inches (f/4.0), with no hole

Franks's natural preference was for the Number 2 mirror, which he expected to perform better due to its greater thickness, promising less flexure, the absence of a central hole, which would mean easier silvering, and its greater weight, which would mean that less counterbalance ballast would be needed at the opposite end of the tube. Nevertheless his first tests were with the Number 1 mirror. Initially these were conducted in the daytime, with the first night-time observations on 1926 Jun 21: "Had two hours with the new telescope from 8.0 to 10.0 PM. Moon tolerable, but, of course, no test. Arcturus and Saturn both gave distorted double images – the ring of [Saturn] unrecognisable. On the other hand the 8½" showed good images of both!" What a disappointment it must have been after all the effort expended!

Over subsequent days, Franks, slackened the mirror retaining ring, rotated the mirror by 90° and changed the flat, but no improvements in the images were forthcoming. Initially he considered that the problem was due to unstable atmospheric seeing and the rapid drop in temperature during the evening. However, on July 8 the seeing was "steady – a perfect night, but the great mirror performed no better….the fault must be with the mirror itself, to produce this optical distortion." Thus on he installed the Number 2 mirror, but "it showed exactly the same kind of distortion as no.1, but smaller" (42). Then on July 15, performing a star test on Arcturus, "to my agreeable surprise, I got a neat image at last! It seems chiefly a matter of atmosphere and temperature?"

Using the telescope during 1926 August and September was a frustrating experience: on some nights acceptable images were obtained, whereas on others observing was impossible. He also tried various aperture masks, without success. Nonetheless, Franks sent the Number 2 mirror to H.N. Irving's of Ipswich for silvering (43). He had to wait until 1927 January before he could use the silvered mirror under excellent conditions, but even then "Mars was mostly distorted, though at times shrunk to a passable disc" (44). Star tests were also disappointing. "It seems pretty conclusive", Franks noted, "that under all the varied tests, it must be the <u>figure</u> of the 24" that is <u>not truly parabolic</u>…The tolerable images are only transitory; never permanent or perfect  - as in all other specula I have seen" (45).

Franks carried out further tests throughout 1927 and into the spring of 1928. It wasn't until 1928 June 1, nearly three years since the telescope was installed, that he was able to conclude "The 24-inch 'Bush' reflector may now be considered as practically in working order". But it was not until a year later that he commenced systematic observations with it (46) and still the views through it were highly variable. On one memorable evening he observed M8, M17 and M20; they were "very fine, much like Lord Rosse's drawing[s] – best views I have had yet" (47).

In accordance with Bush's plans, the telescope was ultimately destined for Nottingham University, so in 1928 August, four months after Bush's death, Franks received a visit from several members of University staff including Dr. Henry H.L.A. Brose (1890-1965; Figure 10). Brose had been a Reader in Physics at the University since 1919 and would later become Professor (48). A few days later a delegation came from the Royal Observatory Greenwich to see the telescope (49), followed the next month by BAA President W.H. Steavenson. Steavenson's main objective was to test a third 24-inch Bush mirror, of even shorter focus, 82-inches at f/3.5!, which he found "satisfactory" (50). It would clearly take some time for the University to make arrangements to receive the telescope.

Franks obtained some good results over the next year, including his best ever views of M42 and M1, but frustratingly these were interspersed with many poor nights. Then during an observing session on 1929 June 28, the driving clock failed and it wasn't repaired until a year later – in the meantime Franks had embarked on another observing project, which was to survey diffuse nebulae with the $6^1/_8$-inch Cooke refractor, and which will be discussed later. By this time he was probably frustrated by the variable results with the Bush reflector, for the next time he recorded observing with it was on 1930 June 30, and six days later he made his final observation with the telescope.

Franks devoted a page of one notebook to summarising his frustrating experiences with the telescope, concluding that he "reluctantly found it <u>unsuitable</u> for <u>visual</u> use. The focus is far too short except for photo. work, for which it was really intended. For critical definition it was nowhere comparable to the $6^1/_8$" Cooke O.G.….The only objects on which it performed satisfactorily were <u>Clusters</u> and <u>Nebulae</u>……Upon these it gave some wonderful views, surpassing anything we had previously seen". He described the mounting as "very rough…causing much waste of time and energy – so that the fatigue involved spoilt any pleasure in using so cumbersome and instrument. Thus for photographic work, "the mounting would have to be scrapped, and a proper modern equipment substituted. Perhaps not worth the effort?"

Clearly the mechanical aspects of the Bush reflector left much to be desired, with Franks noting on one occasion that "the mechanical parts of the 24" gave much trouble, and it requires drastic attention" (51). Mounting and maintaining the figure of such a large mirror is difficult, since it is liable to flex under its own weight. Moreover telescopes of this aperture are much more susceptible to the effects temperature variations and to local seeing conditions. This might explain why good views were only obtained on a few nights. But there were undoubtedly other factors contributing to the disappointing results. Firstly the collimation of such short focal length systems is notoriously critical, as any modern amateur astronomer using a fast Dobsonian telescope will attest, as is maintaining collimation. Franks only mentions collimating the mirrors occasionally, although he might have done it more often. The second factor is the eyepieces that Franks had to hand. These were mostly "simple" designs of the Huyghenian and Ramsden configuration, which, whilst giving good images with long focal length refractors (and some were "borrowed" from the $6^1/_8$-inch Cooke), they are doomed to failure when used with "fast" optics. Later Franks ordered some "Orthoscopic" eyepieces from Cooke, Troughton & Simms, which were slightly better, although one of these he had to send back because it did not meet his expectations. Today we are fortunate in having a great variety of advanced eyepieces available which are specifically designed for fast instruments.

Although located at Brockhurst, the 24-inch reflector was the personal property of Bush, or, upon his death Bush's estate (52). Thus, in spite of Franks's rather damning impression of the telescope, it was finally dismantled in 1934 and transported to Nottingham University on 1935 April 27. Perhaps the University would have more resources available to bring to it. Thus it was set up on the University grounds in Shakespeare Street during the summer, but the staff was never able to get the driving clock working properly, something which Franks himself had struggled with over the year. The instrument was therefore dismantled and put into storage in a cellar. The University building and the telescope were damaged by a German bomb during the "Nottingham Blitz" of 1941 May. What remains of the telescope, including the mirror, is today stored at the University's Highfield campus.

**Franks and the nebulae**

After a lifetime of stellar and planetary observations, it is perhaps surprising that Franks's last two major projects were on the nebulae. Fr. Hagen, with whom Franks corresponded for many years and who had encouraged his star colour observations, suggested that Franks might observe some of the neglected nebulae from William Herschel's catalogues, the existence of some of which was disputed.

Franks embarked on the project during 1927 October, concluding it the following summer. He used the $6^1/_8$-inch Cooke refractor and probably saw this as a welcome diversion from the trials and tribulations of the 24-inch reflector; presumably with the greater light gathering power of the latter, he would have used it to observe the faint nebulae if it had been in full working order. As with his previous projects, Franks went about tracking down the nebulae with gusto, his notebook brimming with observations. The result was a description of 44 objects which were presented in a paper in *Popular Astronomy* (53).

Franks's last major project was also on the nebulae, this time a visual survey of the dark nebulae, stimulated by the recent appearance of E.E. Barnard's *Photographic Atlas of Selected Regions of the Milky Way* (54). The project, again encouraged by Fr. Hagen, was carried out in late 1929. The last observation was made on 1929 December 1, of B 227 in Orion, and his manuscript of the paper despatched on the same day. Out of Barnard's 346 catalogued dark nebulae, he was able to locate about 120 using the Cooke refractor and the results were published the following year in *Monthly Notices of the RAS*, this being his final publication (55).

During the project Franks also spent some time searching for "Baxendell's Nebula", a large and faint nebula observed by Joseph Baxendell (1815-1887) with his 6-inch refractor at Birkdale in 1880 September (56). Baxendell reported that it was located near M2 in Aquarius, about 75' by 52' in size, and similar in character to the nebulosity in the Pleiades, but fainter. The object was listed as NGC 7088 in the *New General Catalogue* and was seen over the years by a number of other well-known observers including J.L.E. Dreyer, who found it without difficulty in a 10-inch refractor, Guillaume Bigourdan, Fr. Hagen, and Max Wolf of Heidelberg. Hagen drew Franks's, and others, attention to the nebula in a note he wrote in *Astronomiche Nachrichten*, encourage observations. Initially Franks had trouble locating the nebula, then on 1929 October 6 he found it: "On careful attention there does seem to be a slight difference in density between the neb[ula] and surrounding starry zones". Fr. Hagen reported Franks's positive observation along with another one made by Fr. O'Connor of Stonyhurst College with the 15-inch (38 cm) refractor. Whilst these visual confirmations were all well and good, the embarrassing truth was emerging that long exposure photographs of the region showed no nebulosity whatsoever, earning it the sobriquet *Baxendell's Unphotographable Nebula*. Hagen argued that it was impossible for all the visual observers to be mistaken and therefore the problem must lie with the photographic process (57). It is now accepted that the object isn't real and that visual sightings of it were due reflections of the nearby bright globular cluster M2.

**The final years**

Although Franks continued to observe after the completion of his work on nebulae, it was much less systematic than previously as the frailties of increasing age became manifest. Working alone was becoming more of a challenge, but he did receive regular assistance in the observatory from a person referred to on the notebooks as "Hubert", especially when the 24-inch reflector was used (58). Most of the notebook entries between 1930 and 1935 were in connexion with visitors to the observatory, meteorological events and other news items, such as the dismantling of the Bush telescope described previously. He also gave talks to various groups in the East Grinstead area.

Franks used to cycle from his home to Brockhurst every day and was a familiar figure in the district (59). In 1933 September, whilst walking, he was knocked down by a cyclist, resulting in a lacerated left arm and a bruised right arm and shoulder. This put him out of action for a month. The following year, 1934 October, he was again taken ill and laid up for another month (60).

Franks's last notebook entry was on 1935 June 7, recording an exceptionally heavy rainfall (61). A few days later a cycling accident befell him from which he never recovered, passing away on June 19 in his eighty-fifth year. It was shortly after this, of course, that Patrick Moore was invited to take change of showing visitors to Brockhurst objects through the telescope.

**Acknowledgements**

I would like to thank Richard Baum for his great kindness and generosity in passing Franks's notebooks to me and for encouraging this research. Richard Pearson has been most generous in allowing me access to his important research on T.W. Bush, as well as giving me permission to reproduce some of the photographs of the Brockhurst Observatory and the Bush telescope. Martin Mobberley and Denis Buczynski provided helpful information about Franks's work and notebooks, including scans of some of these documents. I am grateful to Sue Coppin of the University of Adelaide Archives for permission to use the photograph of Henry Brose. Officers at the Manuscripts and Special Collections at The University of Nottingham were helpful in providing details of items in their archives concerning Franks.

11. Sir George Truscott (1857-1941) of the Baronetcy of Oakleigh, East Grinstead, was Chairman of the printers and stationers, Brown, Knight & Truscott. He was Lord Mayor of London in 1908-9. He was accompanied by his daughter on 1924 July 11.

12. Lady Mary Thomasina Lockyer (1852-953) and her daughter visited on 1926 July 13.

13. Also visiting that night were "Dr. Marshall and son".

14. In 1909-10 Franks wrote three papers on star colours in which he attempted to correlate observed colours with their spectral type: MNRAS, 70, 187-194 (1909); MNRAS, 70, 521 (1910); AN, 185, 97-100 (1910). Whilst these papers were not based on observations made at Brockhurst, he was clearly still very interested in the subject of star colours.

15. 1913 January 18. "Venus. Terminator slightly concave".

16. Notes from 1912 October 9 and 1913 January 31.

17. Franks W.S., notebook, 1931 September 17.

18. Franks W.S., MNRAS, 74, 517-534 (1914).

19. Franks W.S., MNRAS, 75, 96-102 (1914).

20. Franks W.S., MNRAS, 76, 24-36 (1915); MNRAS, 77, 46-54 (1916); MNRAS, 77, 54-64 (1916); MNRAS, 78, 82-90 (1917); MNRAS, 79, 83-98 (1918); MNRAS, 80, 215-227 (1919); MNRAS, 81, 150-154 (1920).

21. The award was made in person by RAS President, J.L.E. Dreyer (1852-1926): Dreyer J.L.E., MNRAS, 83, 428,-430 (1923). Franks notes the event in his notebook entry for 1923 June 8. "RAS meeting. Fr[anks] presented with the Jackson-Gwilt Medal for the micrometrical measures of 2268 stars (1913-1920) published in the 'Monthly Notices'; and star colour obsns. of about 6500 stars".

22. Franks sent Aitken his double star measurements, made between 1915 January and May, on 1915 May 11. Franks W.S., notebook, 1915 May 11.

23. Franks W.S., MNRAS, 75, 37-38 (1914).

24. Franks W.S., notebook, 1914 October 21. The history of the Cooke refractor before Franks was involved with it in 1909 is not known. Frank's note suggests even then, 1914, it was quite old. The author would be delighted to hear from anyone who can shed light on the matter.

25. For example he used the Greenwich instrument to measure OΣ 137 in 1914: Franks W.S., MNRAS, 74, 655-666 (1914).

26. Observations were suspended for more than a month in the summer of 1916 during roof repairs at the observatory. These were eventually "completed after much vexatious delay (War Office Permit etc. During Great War)". Franks W.S., notebook, 1916 September 27. He also found that repairs of the micrometer by T. Cooke & Sons took longer during the war: normally refitting the hairs would take three days, whereas at the end of 1917 it took several months.

58. Investigations have so far not yielded more information about Hubert.

59. Patrick Moore describes in reference 1 how Franks generally wore a distinctive skull cap. He also stated that his beard and diminutive stature had the effect of making him resemble a gnome - in fact the same allusion was mentioned by W.H. Steavenson in his MNRAS obituary of Franks.

60. In his notebook entry for 1935 October 24, Franks describes his illness as "erythema", which presents as a red rash to the skin. It can be triggered by a number of different things, including infection and allergy.

61. Franks recorded a number of unusual weather events of 1935 May. The Silver Jubilee of King George V on May 6 was very warm, at 72.5°F (22.5° Celsius), but the night of May 16/17 brought a frost to Brockhurst, resulting in much damage in the garden. Snow fell in Devon and Cornwall.

62. Many of Franks's drawings from his sketchbook were published in Patrick Moore's paper cited in reference 1.

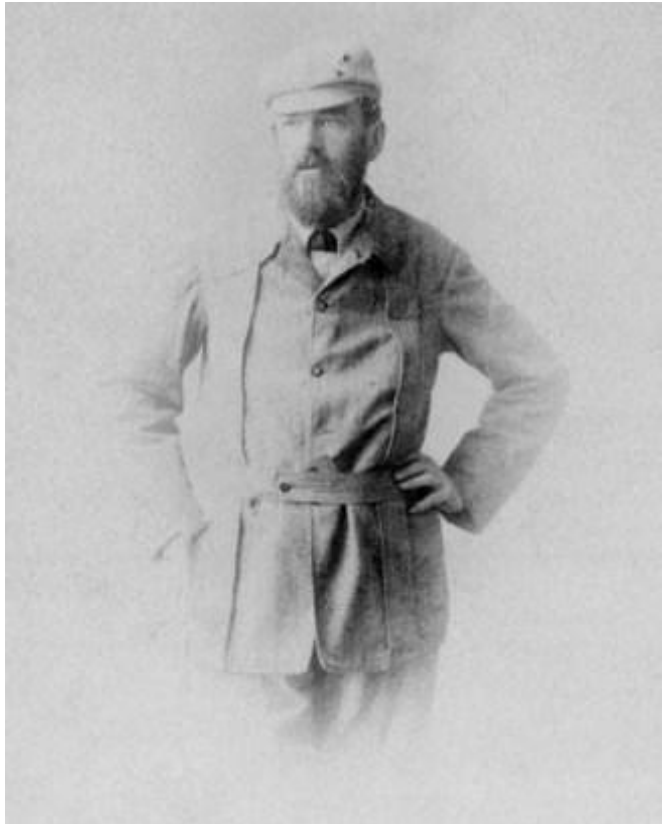

Figure 1: John Franklin-Adams (1843-1912)

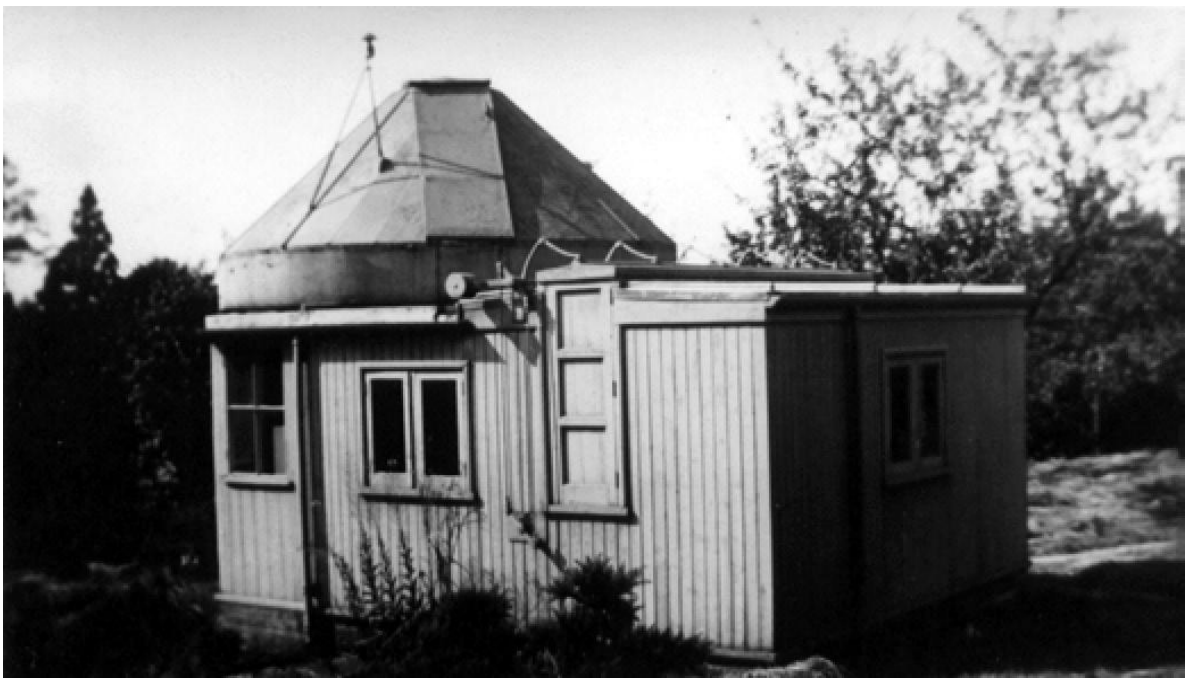

Figure 2: The Brockhurst Observatory, ca. 1932

Note the transit room on the right was added in 1913. The photograph was given to Richard Pearson by Patrick Moore

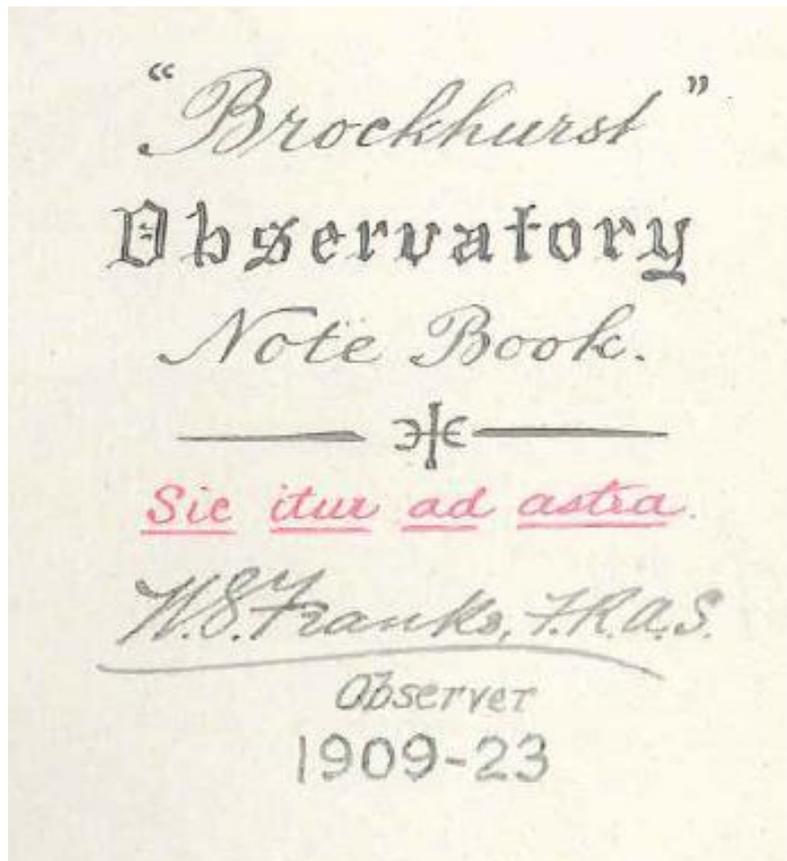

Figure 3: Title page of Franks's first notebook, *Astronomical Diary and Observations (I)*

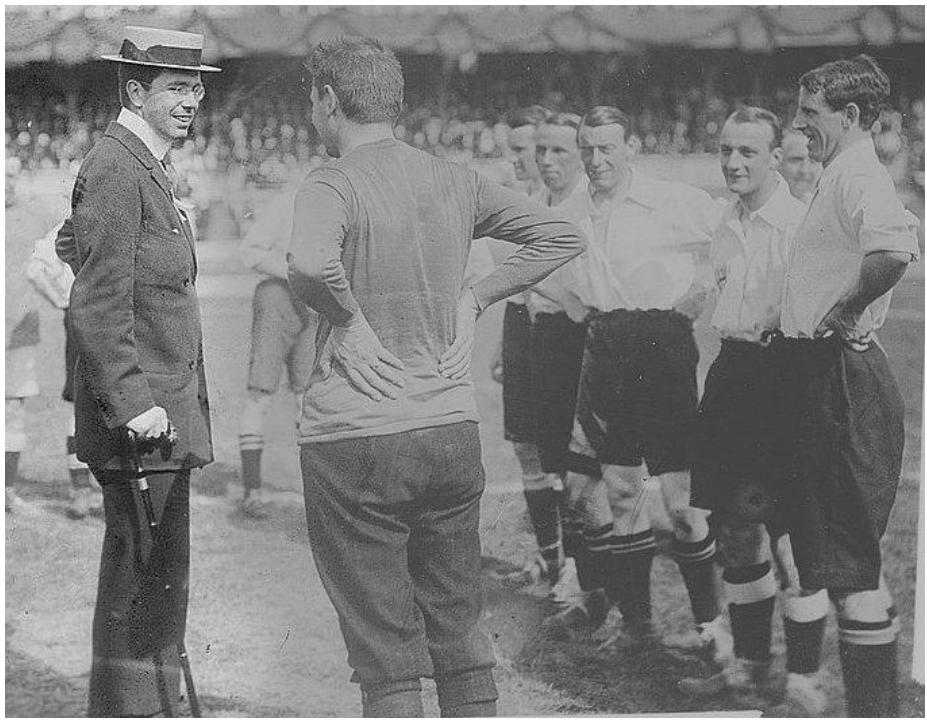

Figure 4: HRH the Crown Prince Gustav Adolf of Sweden talking to English footballers, ca. 1910-15

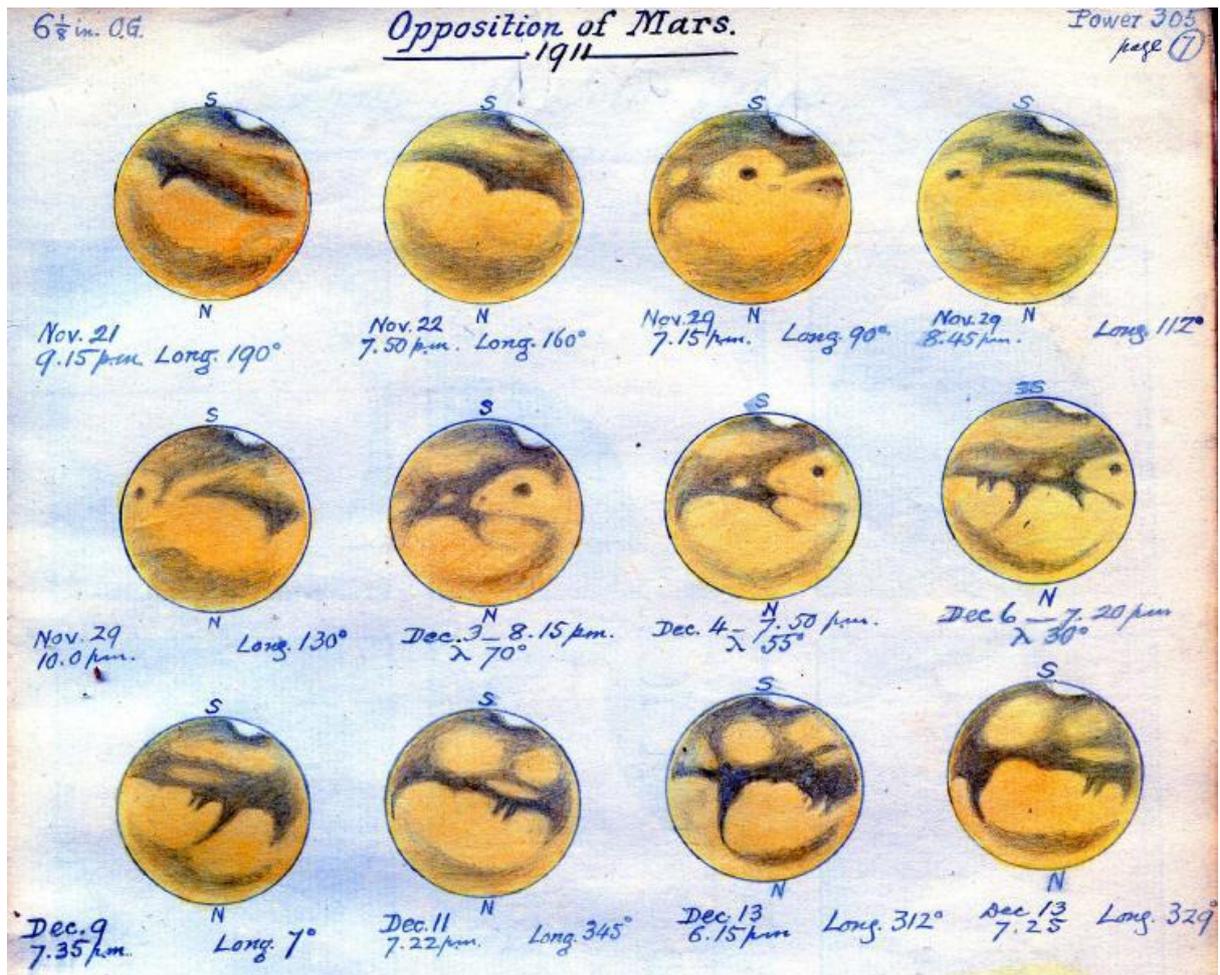

Figure 5: Mars during the 1911 opposition

From Franks's notebook "Astronomical Sketches" (62)

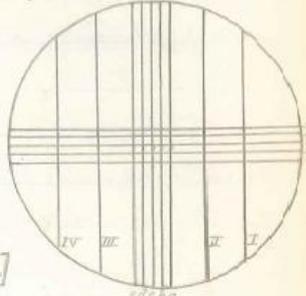

*Particulars of Transit Instrument, erected 4 July, 1913.*

O.G. 2¾ ins. aperture; 32 ins. focus – by Troughton & Simms, London.

Reducing caps for solar observation – 2⅜ in. glass plane

Eyepieces Ramsden (1) power 62; do (2) = power; diagonal prism; Sun-shade
Field 45' (60)

Circle (altitude) 10 ins. diam, divided on silver to 20'; vernier reads to 30."

System of wires – as in annexed diagram:—
a – b = 7½
b – c = 8¼  } mean equatorial interval = 7·5ˢ
c – d = 7½
d – e = 7¼

Striding level – 40 divs. from zero; 1 div. = 1".

Axial illumination – bright field, with ruby screen
2 oil lamps, on Boxwood supports;
allowing for reversal [8 c.p. electric]

Stand cast iron, in 1 piece; azimuth & level screws; duplicate tangent screws.

Case to contain telescope & all accessories; except cast-iron stand.

Co-lat = 38°52'33". Long. 0°0'34" E = 2·27ˢ before Greenwich.

Sidereal Clock by Chas. Frodsham, no 860, in mahogany case (J.F.A)
double-cylinder Invar compensation

Meridian mark.          General View.

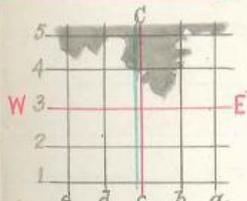

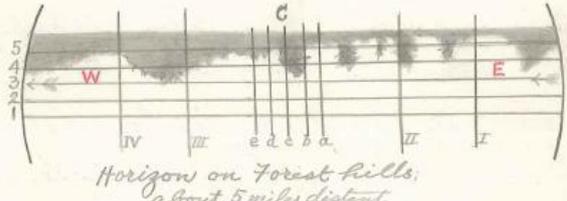

Horizon on Forest hills;
about 5 miles distant

Levelling Screws:– Top nut turned to Left raises axis,  } the other nut
Bottom  "     "     " Right lowers  "  } being slackened.

Azimuth Screws pull towards operator. Tangent Screws push away.

Figure 6: Details of the 2¾-inch Troughton & Simms transit scope at Brockhurst from Frank's notebook

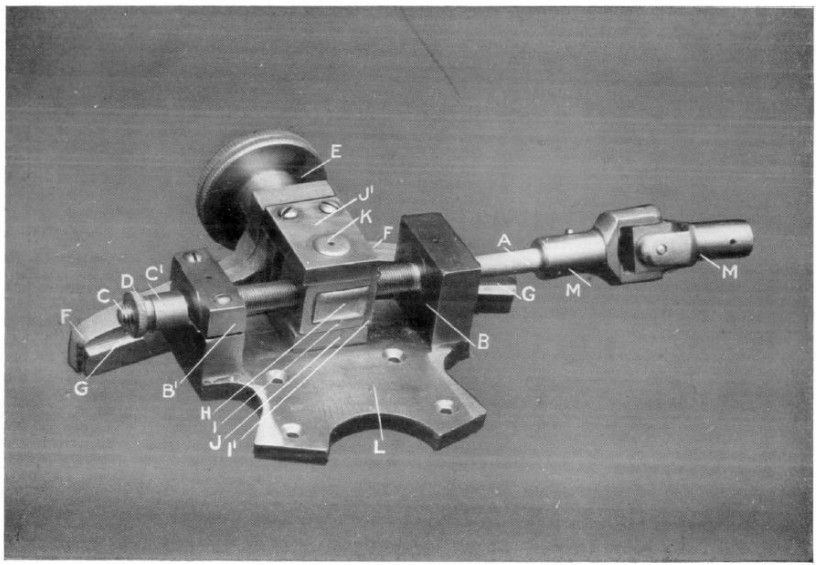

SLOW-MOTION GEAR.

A. Screw shaft.
B. Coned bearing.
B'. Adjustable bearing.
C. End thrust screw, working in boss C'.
D. Lock nut.
E. Clamp head.
FF. Outer clamp arc.
GG. Inner clamp arc.

H. Split nut, sliding in I.
I. Rectangular casing.
I'. Adjustable cover for casing.
J. Lower pivot plate.
J'. Upper adjustable pivot plate.
K. Upper coned pivot.
L. Base plate, carried by arm on polar axis.
MM. Hooke's joint to slow-motion handle.

Figure 7: The new drive gear for the 6 $^1/_8$-inch refractor designed and built in the Brockhurst workshops in 1914. Image from reference (23)

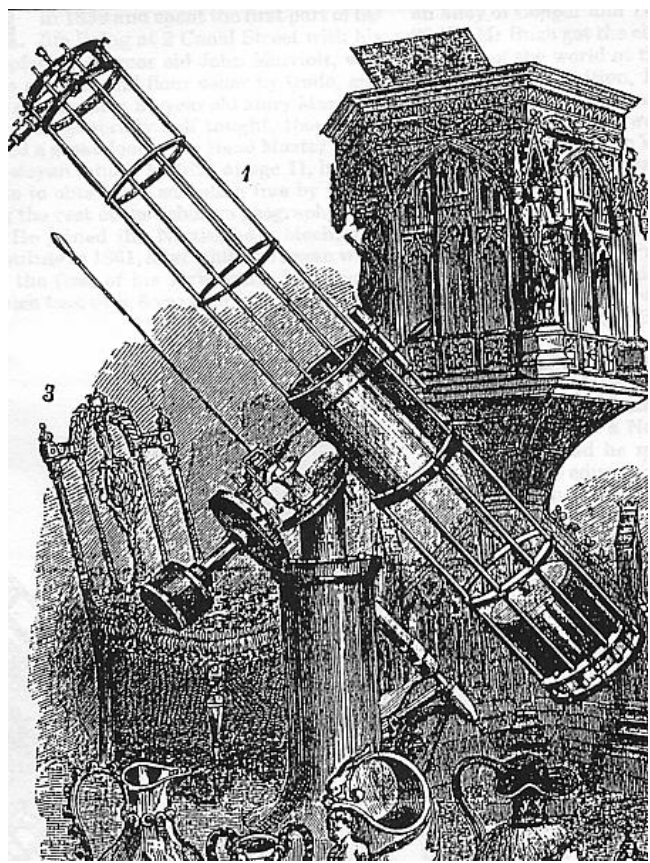

Figure 8: T.W. Bush's 13-inch reflector
Image courtesy of Richard Pearson

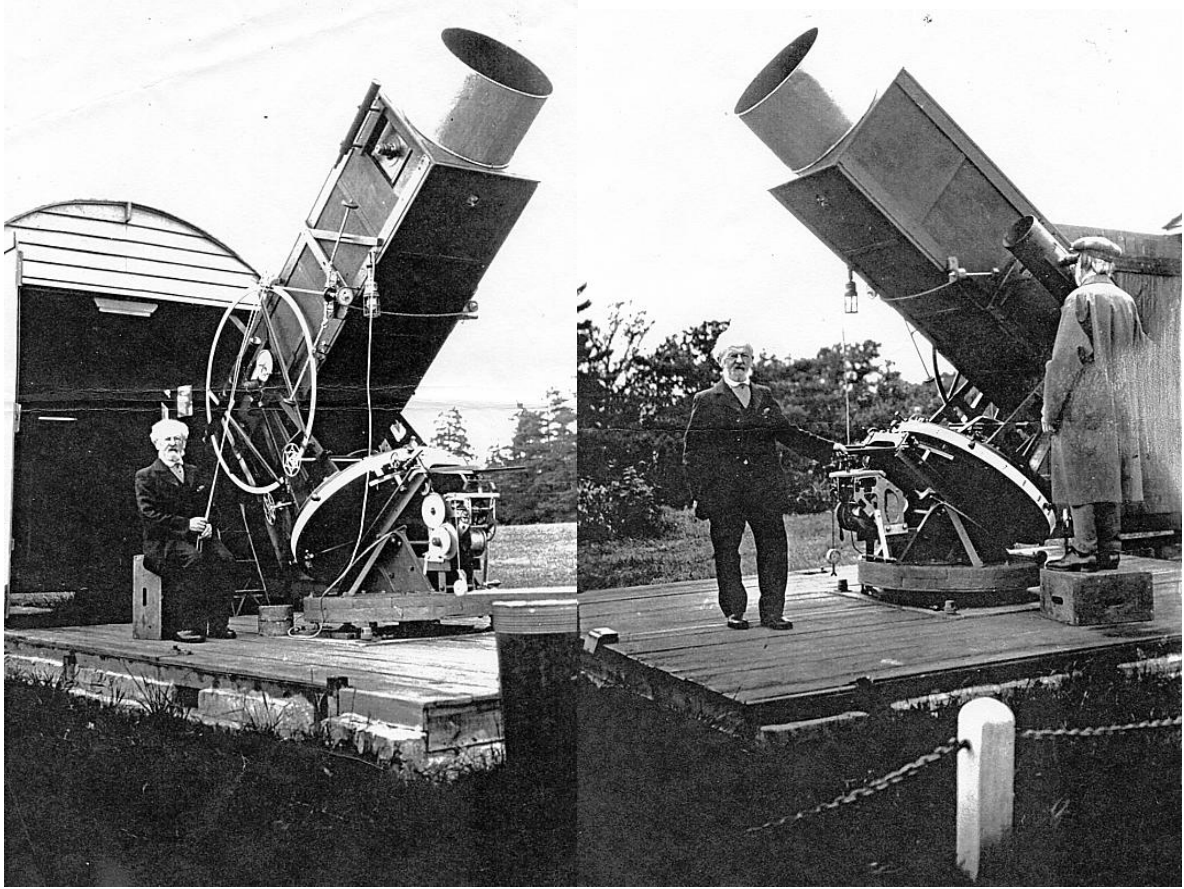

Figure 9: The 24-inch Bush reflector at Brockhurst. (a) W.S. Franks seated, (b) W.S. Franks standing next to the RA axis and T.W. Bush looking into the guide scope eyepiece
The photographs were given to Richard Pearson by Patrick Moore

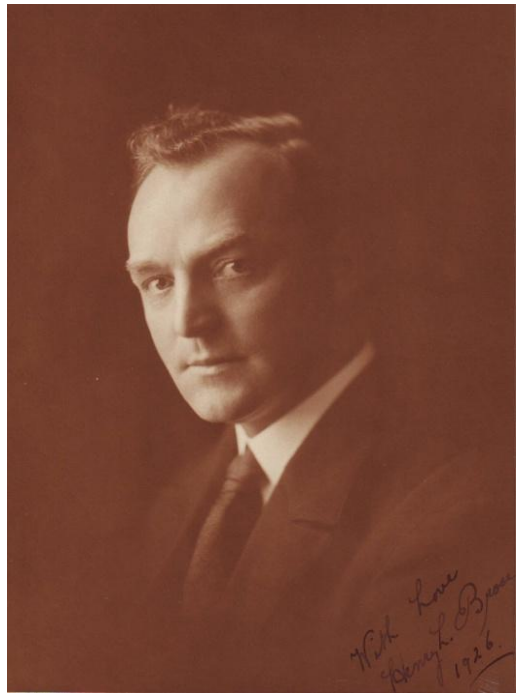

Figure 10: Henry L. Brose in 1926
Image courtesy of the Maude Puddy Collection of the University of Adelaide Archives